\documentclass[11pt, conference, compsocconf, onecolumn]{IEEEtran}
\ifCLASSINFOpdf
\else
\fi
\usepackage{graphicx}
\usepackage{epstopdf}

\begin{document}
\title{Analyzing Medium Access Techniques in\\ Wireless Body Area Network}

\author{N. Javaid, I.Israr, M. A. Khan, A. Javaid, S. H. Bouk, Z. A. Khan$^{\$}$\\
        COMSATS Institute of Information Technology, Islamabad, Pakistan. \\
        $^{\$}$Faculty of Engineering, Dalhousie University, Halifax, Canada.\\
     }

\maketitle

\begin{abstract}
This paper presents comparison of Access Techniques used in Medium Access Control (MAC) protocol for Wireless Body Area Networks (WBANs). Comparison is performed between Time Division Multiple Access (TDMA), Frequency Division Multiple Access (FDMA), Carrier Sense Multiple Access with Collision Avoidance (CSMA/CA), Pure ALOHA and Slotted ALOHA (S-ALOHA). Performance metrics used for comparison are throughput(T), delay(D) and offered load(G). The main goal for comparison is to show which technique gives highest Throughput and lowest Delay with increase in Load. Energy efficiency is major issue in WBAN that is why there is need to know which technique performs best for energy conservation and also gives minimum delay. Simulations are performed for different scenarios and results are compared for all techniques. We suggest TDMA as best technique to be used in MAC protocol for WBANs due to its high throughput and minimum delay with increase in load. MATLAB is the tool that is used for simulation.

\end{abstract}


\section{Introduction}

Energy efficiency is an important issue in WBANs because sensor nodes damage human body tissue. More importantly sensor nodes connected to body are battery operated devices, they have limited life time. So, MAC protocols of WBANs needs to be energy efficient and supports medical applications. It allows integration of low power intelligent sensor nodes. They are used to stream biological information from human body and transmit it to a control device called coordinator. This procedure is very helpful while monitoring health of a person and in case of emergency providing proper medication. MAC protocol plays an important role in determining the energy efficiency of a protocol in WBANs. Traditional MAC protocols focus on improving throughput and bandwidth efficiency. However, the most important thing is that they lack in energy conserving mechanisms. The main source of energy wastage are idle listening, overhearing and packet overhead. By controlling these energy waste sources, maximizes the network lifetime.
\\
\indent WBANs have many advantages like mobility of patient and independent monitoring of patient. It can work on Wireless Local Area Networks (WLANs), Worldwide Interoperability for Microwave Access (WiMAX) or internet to reliably transmit data to a server which is monitoring health issues. There are some requirements for the MAC protocol design to be used in WBANs. Firstly all of protocols must have high QoS (Quality of Service), it must be reliable, it needs to support different medical applications.
\\
\indent By using different Medium Access Techniques, different low power and energy efficient protocols for MAC are proposed. The most important attributes of WBANs are low power consumption and delay. Different techniques are used with different protocol to control the delay and to improve the efficiency of MAC protocol. Techniques like Energy Efficient low duty cycle MAC protocol [5],  Energy Efficient TDMA based MAC protocol [6], Traffic Adaptive MAC protocol [9] are used to improve energy efficiency and to control delay.
\\
\indent The important techniques of MAC protocol for WBANs are Time Division Multiple Access (TDMA) and Carrier Sense Multiple Access with Collision Avoidance (CSMA/CA). Frequency Division Multiple Access (FDMA) is very close to TDMA however it is not used due to complexity in its hardware. Pure Aloha and SLOTTED Aloha are not used due to collision problems and high packet drop rates as well as low energy efficiency. There are several challenges in realization of the perfect Multiple Access Technique for MAC protocol design. A comprehensive overview to these accessing techniques is shown in table I.
\section{Related Work And Motivation}
\indent In paper \cite{1}, authors propose using a wake-up radio mechanism MAC protocol for wireless body area network. Comparison of TDMA with CSMA/CA is also done in this paper. Proposed MAC protocol save energy by node going to sleep when there is no data and can be waked up on-demand by wake-up radio mechanism. This protocol works on principle of on-demand data. It reduces the idle time consumption of a node to a great extent. However, emergency traffic are not discussed in this paper, which is a major issue in WBANs.\\

\indent PHY and MAC layers of IEEE 802.15.6 standard are discussed by author in paper \cite{8}. They stated specifications and identified key aspects in both layers. Moreover bandwidth efficiency with increase in payload size is also analyzed. They also discuss the different modes of security in the standard. However, bandwidth efficiency of the standard is only investigated for CSMA/CA. Also they not discussed throughput and delay.\\

\indent Authors in \cite{3} state that the IEEE 802.15.4 standard is designed as a low power and low data rate protocol with high reliability. They analyze unslotted version of protocol with maximum throughput and minimum delay. The main purpose of 802.15.4 is to give low power, low cost and reliability. This standard defines a physical layer and a MAC sub layer. It operates in either beacon enabled or non beacon mode. Physical layer specifies three different frequency ranges: 2.4 GHz band with 16 channels, 915 MHz with 10 channels and 868 MHz with 1 channel. Calculations are performed by considering only beacon enabled mode and with only one sender and receiver. However, it is high power consumed standard. As number of sender increases, efficiency of 802.15.4 decreases. Throughput of 802.15.4 declines and delay increases when multiple radios are used because of increase in number of collisions.\\

\indent In paper \cite{9}, author proposes a modified MAC protocol for WBAN which focuses on simplicity, dependability and power efficiency. It is used in Contention Access period and CSMA/CA is used in contention free period. Data is transmitted in the contention free period where as CAP is only used for Command packets and best effort data packets. However, propagation delay is not neglected which we consider in our comparison and also interference from other WBAN nodes are not taken into account while doing calculation. Technique which are used by the author have high delay as compared to TDMA and FDMA.\\

\indent In this paper \cite{12}, authors introduced a TDMA-based energy efficient MAC protocol for in-vivo communications between mobile nodes in BSNs using uplink/downlink asymmetric network architecture. They also proposed TDMA scheduling scheme and changeable frame formats. The latency optimization is discussed and the performance is improved by reducing the data slot duration. However they have not elaborated about throughput and delay sensitive application.\\

\indent Energy Efficient TDMA based MAC Protocol is described in \cite{2}. Protocol in this paper minimizes the amount of idle listening by sleep mode this is to reduce extra cost for synchronization. It listens for synchronization messages after a number of time frames which results in extremely low communication power. However, this protocol lacks wake-up radio mechanism for on demand traffic and emergency traffic.\\

\indent Authors describe energy efficient low duty cycle MAC protocol for WBANs in paper \cite{5}. TDMA are compared with CSMA/CA. TDMA based protocol outperforms CSMA/CA in all areas. Collision free transfer, robustness to communication errors, energy efficiency and real time patient monitoring are the flaws that are overcome in this paper. However, synchronization is required while using TDMA technique. With increase in data, TDMA energy efficiency decreases due to queuing. As network topology changes TDMA experiences degradation in performance.\\

\indent In this \cite{11}, authors propose a new protocol MedMAC and they elaborate novel synchronization mechanism, which facilitates contention free TDMA channels, without a prohibitive synchronization overhead. They focus on power efficiency of MedMAC. Also they show that MedMAC performs better than IEEE 802.15.4 for very low data rate applications, like pulse and temperature sensors (less than 20 bps). However, they have discussed about collisions but they have not focused on delay in the applications.\\

\indent In paper \cite{7}, authors propose technique for mechanism of low power for WBAN, that defines traffic patterns of sensor nodes to ensure power efficient and reliable communication. They classify traffic pattern into three different traffic patterns (normal traffic, on-demand traffic and emergency traffic) for both on-body and in-body sensor networks. However they have not taken care for the delay and throughput. Also complete implementation of their proposed protocol is still to be done.\\

\indent An Ultra Low Power and Traffic adaptive protocol designed for WBANs is discussed in \cite{4}. They used a traffic adaptive mechanism to accommodate on-demand and emergency traffic through wake-up radio. Wake-up radio is low power consumption technique because it uses separate control channel with data channel. Comparison of power consumption and delay of TA-MAC with IEEE 802.15.4, Wise MAC, SMAC are done in this paper.\\

\indent Authors evaluate performance of IEEE 802.15.4 MAC, Wise MAC, and SMAC protocols for a non-invasive WBAN in terms of energy consumption and delay in \cite{10}. IEEE 802.15.4 MAC protocol are improved for low-rate applications by controlling the beacon rate. In addition, beacons are sent according to the wakeup table maintained by the coordinator. However, authors have not discussed delay and offered load in their paper.\\

\indent In this paper \cite{6}, authors introduce a context aware MAC protocol which switch between normal state and emergency state resulting in dynamic change in data rate and duty cycle of sensor node to meet the requirement of latency and traffic loads. Also they use TDMA frame structure to save power consumption. Additionally a novel optional synchronization scheme is propose to decrease the overhead caused by traditional TDMA synchronization scheme. However, throughput in this paper is not addressed.\\

\begin{table*}
\caption{Comprehensive Table Of Multiple Access Techniques}
\begin{center}
    \begin{tabular}{| p{2cm} | p{2cm} | p{2cm} | p{2cm} | p{2cm} | p{1.7cm} | p{2cm}| p{1.5cm}|}
    \hline
    Technique & Features & Advantages & Disadvantages & Application & Synchronization Required & Modulation scheme/Technique & Probability Of Collision \\ \hline
    TDMA & Divides radio spectrum in time slots & Flexible bit rate &Wide timing synchronization &Digital and Analog systems  & Yes & DQPSK, GMSK and GFSK &Low \\ \hline
    FDMA & Transmit simultaneously and continuously  & Reduced information bit rate & Precise filtering & Analog systems & Yes & FSK and FM &Low\\ \hline
    CSMA/CA & Carrier sensing with collision avoidance & Avoids data collision  & Inappropriate for large/active networks & 802.15.4 (WPAN) & No & DSSS and FHSS & Intermediate\\ \hline
    Pure ALOHA & Sends data without sensing medium & Adaptive to varying number of stations & Requires queuing buffers for retransmission & Ethernet standard based on the ALOHA network/UMTS  & No & N/A &Very High\\ \hline
    S-ALOHA & Divided into time slots & Doubles the efficiency of ALOHA & Synchronization and queuing buffers required & It is used in different frequencies with the same radio front-end & No & GMSK &High\\ \hline
    \end{tabular}
\end{center}
\end{table*}

\section{Introduction Of Multiple Access Techniques}
Channel access mechanisms provided by Medium Access Control(MAC) layer are also expressed as multiple access techniques. This made it possible for several stations connected to the same
physical medium to share it. Multiple access Techniques have been used in different type of networks. Each technique is used according to its requirement.
In this paper, we are comparing behavior of different multiple access techniques with change in throughput, delay and offered load.
We have plotted them considering three scenarios.
\\
{(1)} Offered load as a function of delay.
\\
{(2)} Throughput as a function of delay.
\\
{(3)} Offered load as a function of throughput.

\subsection{CSMA/CA}
CSMA/CA is a extended version of CSMA. Collision avoidance is used to enhance performance of CSMA by not allowing node to send data if other nodes are transmitting. In normal CSMA nodes sense the medium if they find it free, then they transmits the packet without noticing that another node is already sending the packet, this results in collision. To improve the probability of collision CSMA/CA was proposed, CSMA/CA results in the improvement of collision probability.\\
\indent It works with principle of node sensing medium, if it finds medium to be free, then it sends packet to receiver. If medium is busy then node goes to back-off time slot for a random period of time and wait for medium to get free. With improve CSMA/CA RTS/CTS exchange technique node send Request to send (RTS) to receiver after sensing the medium and finding it free. After sending RTS, node waits
for Clear To Send (CTS) message from receiver. After message is received, it starts transmission of data, if node does not receive CTS message
then it goes to back-off time and wait for medium to get free. CSMA/CA is a layer 2 access method.  It is used in 802.11 wireless LAN and other wireless communication.
\begin{figure}[!h]
\centering
\caption{Timing diagram of CSMA/CA}
\includegraphics[width=3.5 in, height=2 in]{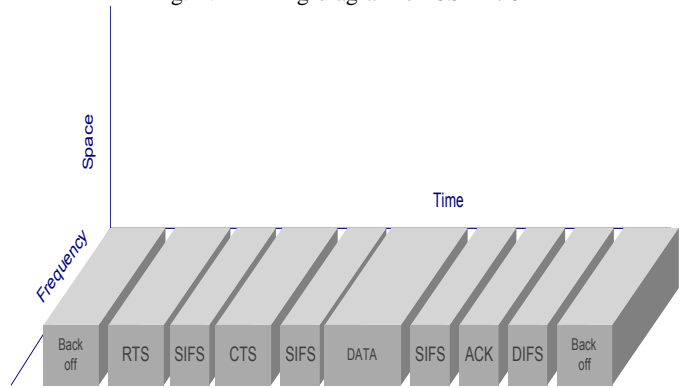}
\end{figure}
\\
Equations which we have used for plotting of CSMA/CA in three scenarios are given below
\\\\
Relation of D and T
\begin{eqnarray}
D=(exp(2*T)-1)*((K-1)/(2+2*a+1)+1+a.
\end{eqnarray}
\\
Relation of D and G
\begin{eqnarray}
D=(G.*(1+G+L.*G.*(1+G+L.*G./2))*exp-G.(*(1+2*L)))./(G.*(1+2*L)-\nonumber\\
(1-exp(-L.*G))+(1+L.*G).*exp(-G.*(1+L)))
\end{eqnarray}
\\
Relation of G and T
\begin{eqnarray}
T=(G.*(1+G+a.*G.*(1+G+a.*G./2)).*exp(-G.*(1+2*a)))./(G.*(1+2  \nonumber\\
*a)-(1-exp(-a.*G))+(1+a.*G). \exp(-G.*(1+a)))
\end{eqnarray}
\\
\begin{figure}[!h]
\centering
\caption{Flow Chart of CSMA/CA}
\includegraphics[width=3.5 in, height=6.5 in]{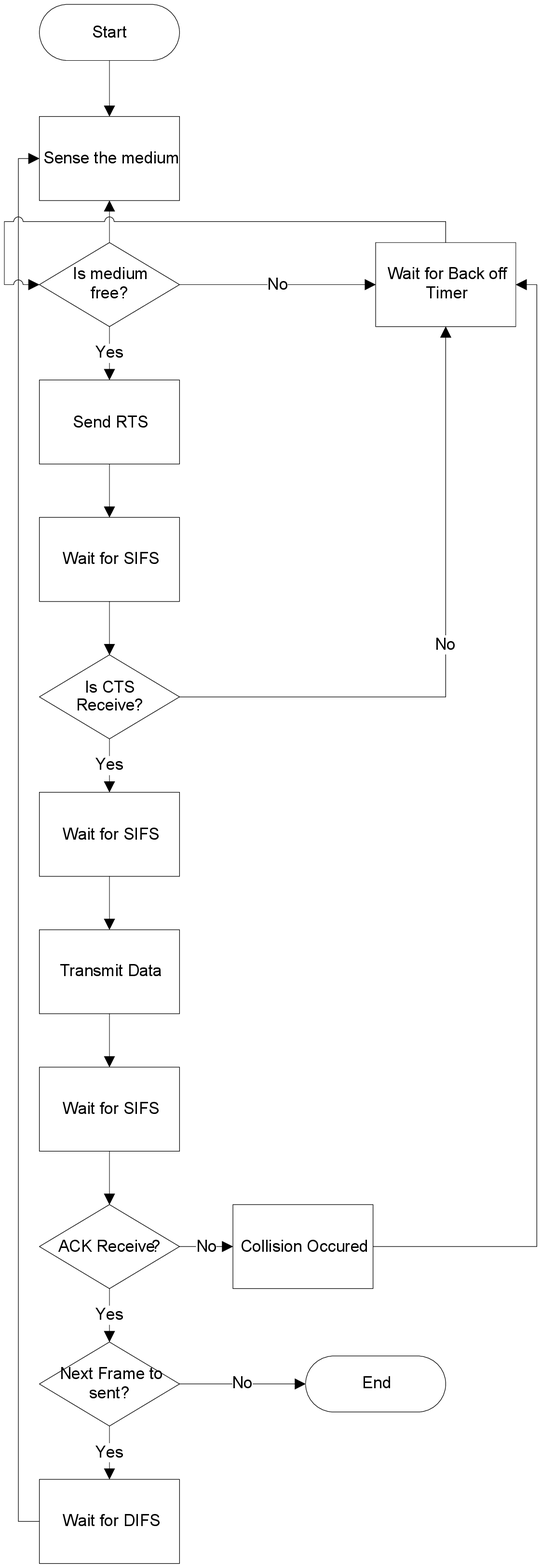}
\end{figure}
Fig 2 shows a flow chart to describe the functionality of CSMA/CA. In CSMA/CA, each node first sense the channel and when channel is free node sends RTS (Request To Send) packet to intended destination and if channel is busy, node goes to back-off timer. After sending RTS packet node waits for SIFS (Short Inter Frame Space) time. If CTS is successfully received then node waits for SIFS time otherwise it goes to back-off time. In back-off time state, node checks for the medium to get free. After SIFS time node start transmitting data packets towards destination node. Then node waits for SIFS and checks for successful reception of ACK (Acknowledgement) packet from destination node. If ACK packet is successfully received, then nodes check for available data packets. However if ACK is not received which results in collision, nodes goes to back-off timer state. If there are no data packets to be sent node terminates communication. However, if data packets are to be sent, node again checks for medium to get free and this process repeats for every data packet.

\subsection{Pure ALOHA}
Pure ALOHA is the first random access technique introduced and it is so simple that its implementation is straight forward. It belongs to the family of contention-based protocols, which do not guarantee the successful transmission in advance. In this whenever a packet is generated, it is transmitted immediately without any further delay. Successful reception of a packet depends only whether it is collided or not with other packets. In case of collision, the collided packets are not received properly. At the end of packet transmission each user knows either its transmission successful or not.
\\
\indent If collision occurs, user schedules its re-transmission to a random time. The randomness is to ensure that same packet do not collide repeatedly. An example of Pure ALOHA is depicted in Fig 3. Each packet is belongs to a separate user due to the fact that population is large.
Fig 4 shows a flow chart for Pure ALOHA and Slotted ALOHA. In ALOHA technique node checks for the availability of data packets to be transmitted.
If they are available then node transmits them otherwise process ends.
\begin{figure}[!h]
\centering
\caption{Pure ALOHA timing diagram}
\includegraphics[width=3 in, height=3 in]{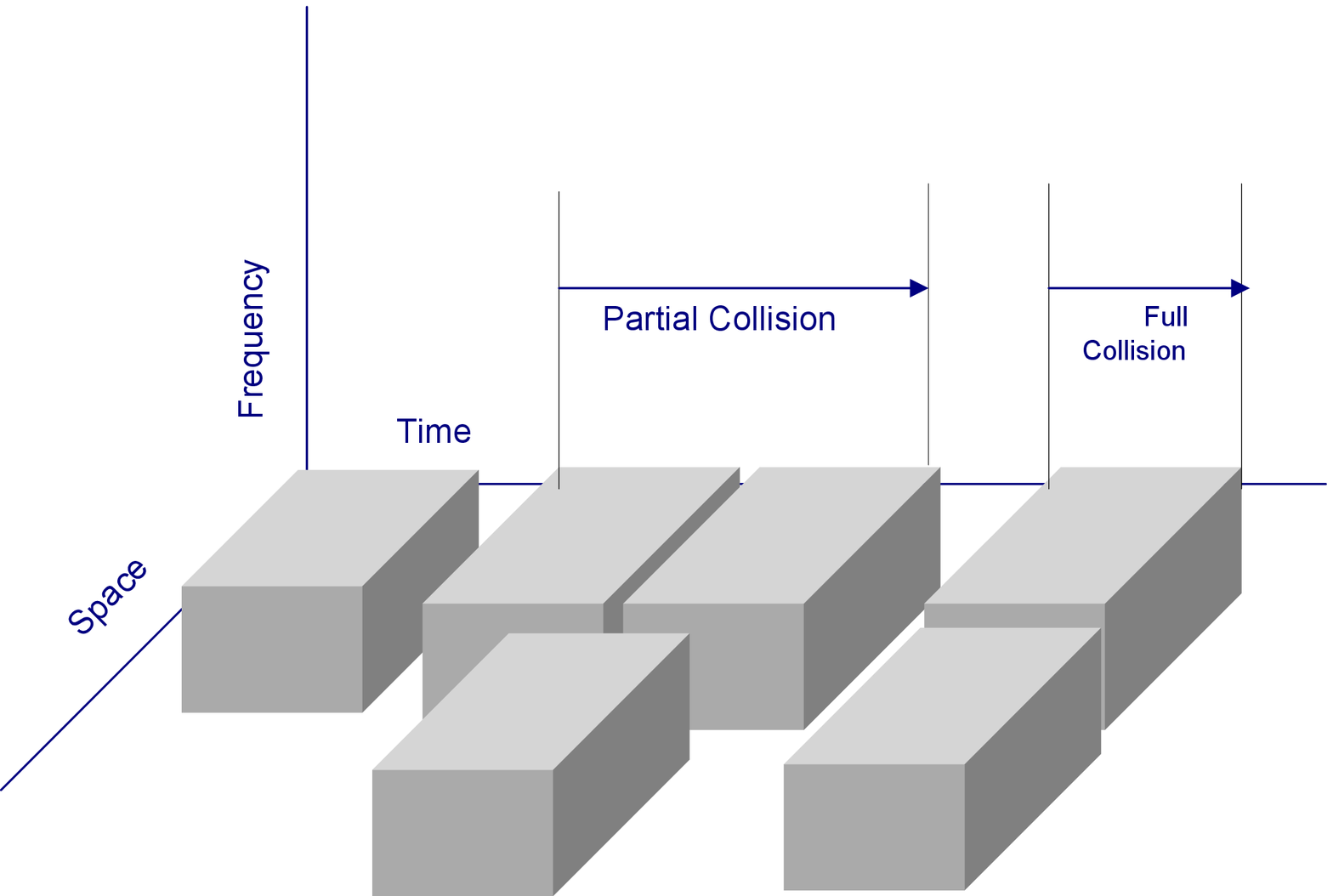}
\end{figure}
\\\\
Relation of T and G
\begin{eqnarray}
T=&G*exp(-2*G)
\end{eqnarray}
Relation of T and D
\begin{eqnarray}
D=&(exp(2*S)-1)*((K-&1)/2+2*a+1)+1+a
\end{eqnarray}
Relation of D and G
\begin{eqnarray}
D=(exp(G)-1*((K-1)/2+2*a+1)+1+a
\end{eqnarray}
\begin{figure}[!h]
\centering
\caption{Flow Chart of Pure ALOHA and S-ALOHA}
\includegraphics[width=3 in, height=3 in]{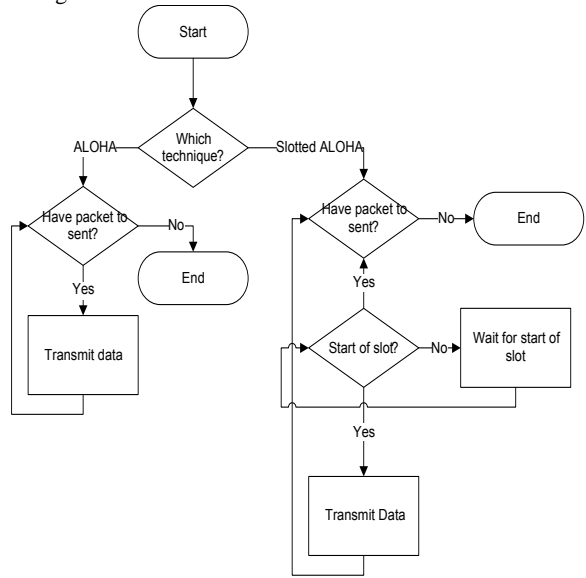}
\end{figure}

\subsection{Slotted ALOHA}
Slotted ALOHA is a variant of Pure ALOHA with channel is divided into slots. Restriction is imposed on users to start transmission on slot boundaries only. Whenever packets collide, they overlap completely instead of partially. So only a fraction of slots in which packet is collided is scheduled for re-transmission. It almost doubles the efficiency of Slotted ALOHA as compared to Pure ALOHA.
Functionality of Slotted ALOHA is shown in Fig 5. Successful transmission depends on the condition that, only one packet is transmitted in each frame. If no packet is transmitted in a slot, then slot is idle. Slotted Aloha requires synchronization between nodes which lead to its disadvantage.
\begin{figure}[!h]
\centering
\caption{Slotted ALOHA timing diagram}
\includegraphics[width=3 in, height=3 in]{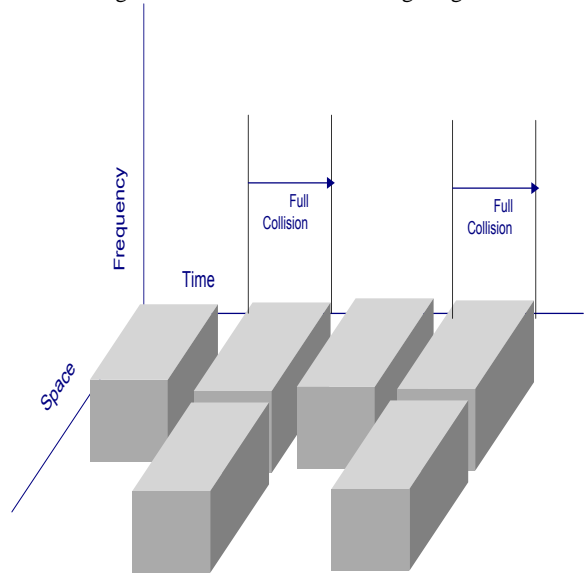}
\end{figure}
Relation of T and G
\begin{eqnarray}
T&=&G*exp(-G)
\end{eqnarray}
Relation of T and D
\begin{eqnarray}
D&=&(exp(S)-1)*((K-1)\nonumber\\
& &/2+2*a+1)+1.5+a
\end{eqnarray}
Relation of D and G
\begin{eqnarray}
D&=&(exp(G)-1)*((K-1)/ \nonumber\\
& &2+2*a+1)+1.5+a
\end{eqnarray}
\\
\indent From Fig 4, in Slotted ALOHA transmission of data packets can occur only at start of slot only. Data packets wait for the start of slot to begin transmission.
If there are no data packets to transmit then process terminates.
\subsection{TDMA}
TDMA works with principle of dividing time frame in dedicated time slots, each node sends data in rapid
succession one after the other in its own time slot. Synchronization is one of the key factors while applying TDMA. It uses full channel width, dividing it
into two alternating time slots. TDMA uses less energy than others due to less collision and no idle listening. TDMA protocols are more power efficient than other multiple access protocols because nodes transmits only in allocated time slots and all the other time in inactive state. A packet generated by node suffer three type of delays as it reaches receiver.
\\
{(1)}Transmission delay.
\\
{(2)}Queuing delay.
\\
{(3)}Propagation delay.
\\
\\
Equations which we have used to plot TDMA in three scenarios are given below:
\begin{figure}[!h]
\centering
\caption{Timing diagram of TDMA}
\includegraphics[width=3 in, height=2 in]{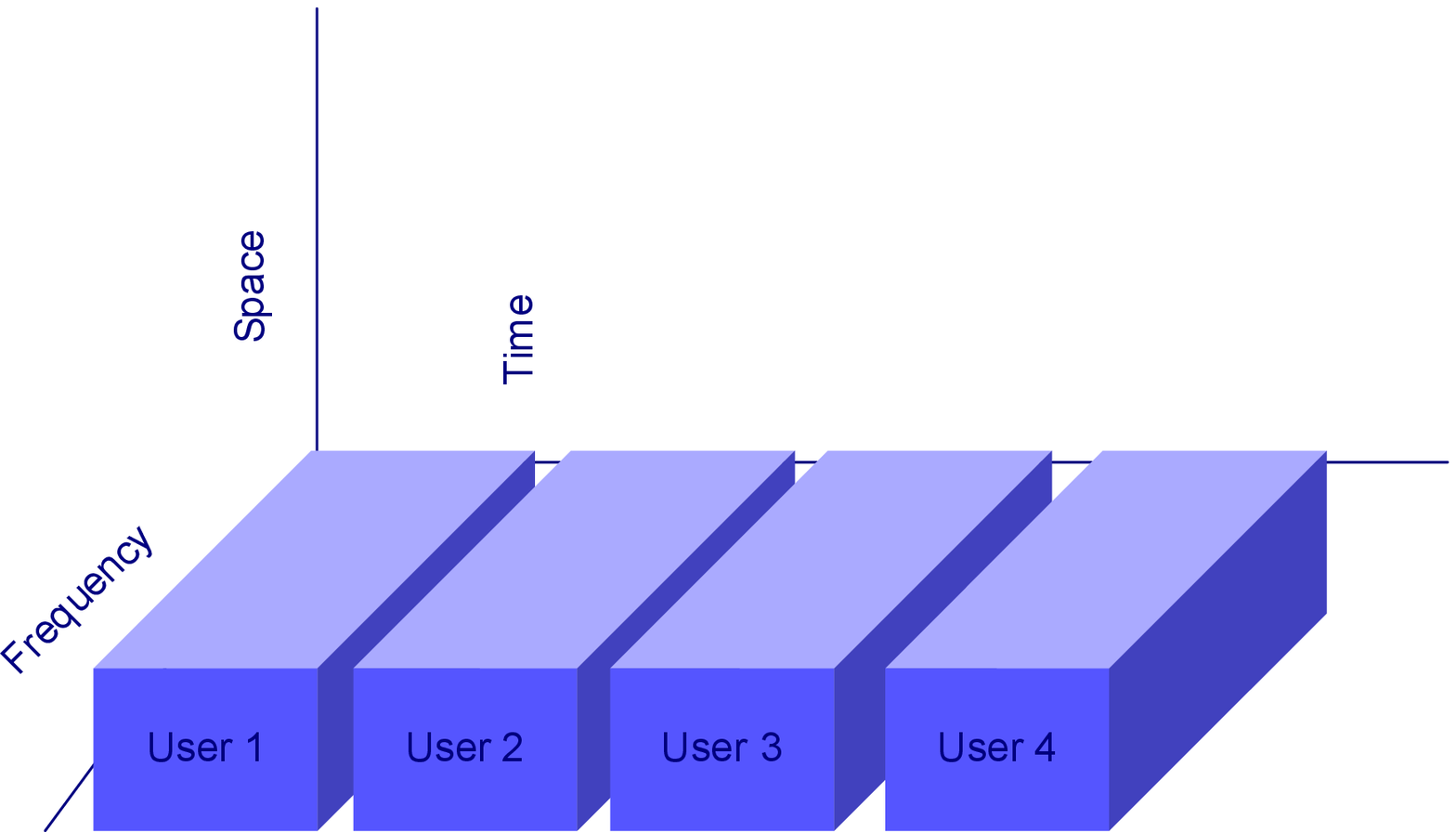}
\end{figure}
\\
Relation of D and T
\begin{eqnarray}
D=L/C+q/(2*(1-q))*N*L/C+N*L/(2*C)
\end{eqnarray}
\\
Relation of G and T
\begin{eqnarray}
T=L/C+G/(2*(1-G))*N*L/C*N*L/(2*C)
\end{eqnarray}
\\
Relation of D and G
\begin{eqnarray}
D=L/a+q./(2*(1-q))*N*L/a+N*L/(2*a)
\end{eqnarray}
\begin{figure}[!h]
\centering
\caption{Flow Chart of FDMA and TDMA}
\includegraphics[width=3 in, height=3 in]{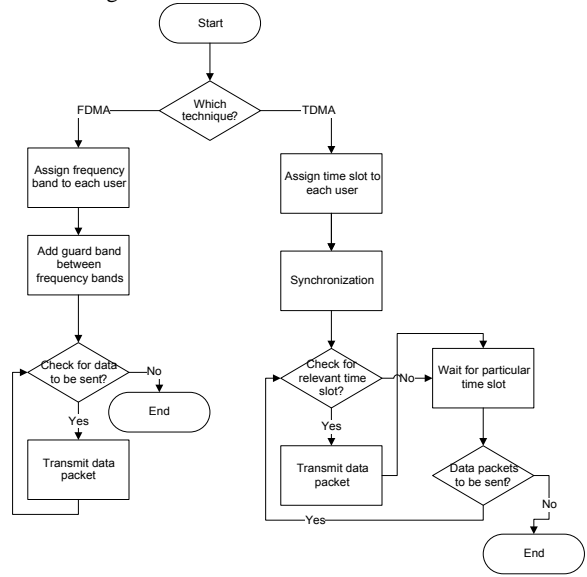}
\end{figure}
\\
\begin{table}
\caption{Description of Parameters used in Equations}
\begin{center}
    \begin{tabular}{ | p{2.5cm} | p{2.5cm} |}
    \hline
    Parameters & Description\\ \hline
    D & Delay\\ \hline
    T & Throughput\\ \hline
    C & Cycle length\\ \hline
    L & Number of packets\\ \hline
    N & Number of nodes\\ \hline
    q & Number of packets in queue\\ \hline
    G & Offered load\\ \hline
    a & Normalized end-to-end delay\\ \hline
    K & Kaapa\\ \hline
    \end{tabular}
\end{center}
\end{table}

Fig 7 shows a flow chart for TDMA. In TDMA, first of all each node is assigned a particular time slot for its transmission. Synchronization is done between source node and destination node. Node checks for its particular time slot and transmits data packets in its relevant time slot otherwise waits for its relevant time slot. If packets are not available for transmission, communication terminates.  Otherwise node checks for availability of slot and this process repeats until communication terminates.

\subsection{FDMA}
FDMA is a basic technology in analog Advanced Mobile Phone Service (AMPS), most widely-installed cellular phone system installed in North America. With FDMA, each channel can be assigned to only one user at a time.
Each node share medium simultaneously though transmits at single frequency. FDMA is used with both analog and digital signals. It requires high-performing filters in radio hardware, in contrast to TDMA and CSMA. As each node is separated by its frequency, minimization of interference between nodes is done by sharp filters. In FDMA a full frame of frequency band is available for communication, In FDMA a continuous flow of data is used, which improves efficiency of sending data. The division of frequency bands among users is shown in Fig 8.
\begin{figure}[!h]
\centering
\caption{Frequency distribution in FDMA}
\includegraphics[width=3 in, height=2 in]{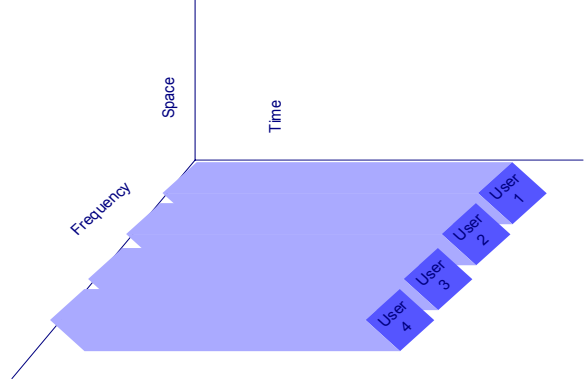}
\end{figure}
Relation of D and G
\begin{eqnarray}
D=N*L/a+q/(2*(1-q))*N*L/a
\end{eqnarray}
Relation of D and T
\begin{eqnarray}
D=N*L/C+q/(2*(1-q))*N*L/C
\end{eqnarray}
Relation of T and G
\begin{eqnarray}
T=N*L/C+G/(2*(1-G))*N*L/C
\end{eqnarray}
From flow chart of FDMA in Fig 7, each node is assigned different frequency band to access the medium. Each frequency band is separated by a guard band which avoids the interference between adjacent nodes/frequency bands. If data is available for transmission, node transmits it without any further process. It terminates communication for unavailability of data packets.

\section{Mathematical Modeling Of Throughput for Multiple Access Techniques}
In this section we are going to calculate the throughput of different multiple access techniques. Data is transferred from sender to receiver using one of the techniques, throughput due to these techniques have been calculated. Due to less difference between sender and receiver, there are no packet losses due to collision, no packets are lost due to buffer overflow. For the calculation of throughput we are assuming a perfect channel. Throughput is calculated for all access techniques through following equation.

\begin{eqnarray}
T&=& \frac{8.x}{delay(D) (x)}
\end{eqnarray}
In equation 16 D is delay, T is throughput and x is the no of bits passing through the frame.
\subsubsection{Throughput of Pure ALOHA}
The calculation for the throughput of ALOHA is done by formula given in equation 16 and the delay which it experience is calculated below
\begin{eqnarray}
D=T_{data} + T_{que}
\end{eqnarray}
\\
Following notations are used
\\\\
$T_{data}=Time$ $for$ $data$ $to$ $reach$ $end$ $of$ $frame$
\\
$T_{que}=Time$ $for$ $queuing$
\\

\subsubsection{Throughput Of TDMA}

Throughput is calculated by using equation 16. Delay which a packet experiences as it reaches from sender to destination is calculated as following

\begin{eqnarray}
D=T_{oh}+T_{ack}+T_{g}+T_{sync}+T_{ta}
\end{eqnarray}
\\
Different time delay given in equation 18 can be calculated by following equations
\\
\begin{eqnarray}
T_{oh} & =&\frac{N_{oh}}{f_{c}}
\\
T_{ack}&=&\frac{N_{ack}}{f_{c}}
\\
T_{sync}&=&\frac{N_{syn}}{f_{c}}
\\
T_{data}&=&\frac{N_{data}}{f_{c}}
\end{eqnarray}

Following notations are used
\\\\
$T_{sync}=Synchronization$ $time$
\\
$T_{data}=Time$ $for$ $data$ $to$ $reach$ $end$ $of$ $frame$
\\
$T_{ta}=Turnaround$ $Time$
\\
$T_{ack}=Acknowledgement$ $time$
\\
$T_{oh}=OverHead$ $time$
\\
$T_{g}=Guard$ $time$
\\
$f_{c}= Communication$ $Data$ $Rate$
\\
$N_{oh}=Total$ $overhead$ $bits$
\\
$N_{ack}=ACK/NACK$ $message$ $bits$
\\
$N_{syn}=Total$ $synchronized$ $bits$
\\
$N_{data}=Total$ $data$ $bits$
\\

\subsubsection{Throughput of S-ALOHA}
The calculation for the throughput of S-ALOHA is done by formula given in equation 16 and the delay which it experience is calculated below
\begin{eqnarray}
D=T_{ack}+T_{syn}+T_{ta}+T_{idle}+ T_{bon}
\end{eqnarray}
\\
Different Time delay given in equation 23 can be calculated by following equations
\begin{eqnarray}
T_{ack}&=&\frac{N_{ack}}{f_{c}}
\\
T_{sync}&=&\frac{N_{sync}}{f_{c}}
\end{eqnarray}

Following notations are used
\\\\
$T_{bon}=Time$ $for$ $data$ $to$ $be$ $transmitted$ $at$ $slot$ $boundaries$
\\
$T_{idle}=Idle$ $time$ $after$ $a$ $transmission$
\\
$T_{ta}=Turnaround$ $Time$
\\
$T_{ack}=Acknowledgement$ $time$
\\
$N_{syn}=Total$ $synchronized$ $bits$
\\
$f_{c}= Communication$ $Data$ $Rate$
\\
$N_{ack}=ACK/NACK$ $message$ $bits$
\\
\subsubsection{Throughput of FDMA}
Throughput of FDMA is very close to TDMA. There is very little difference between throughput of the two multiple access techniques. The calculation for the throughput of FDMA is calculated by formula given in equation 16 and the delay which it experience is calculated below
\begin{eqnarray}
D=T_{oh}+T_{ack}+T_{g}+T_{ta}+T_{data}
\end{eqnarray}
\\
Different time delay given in equation 26 can be calculated by following equations
\\
\begin{eqnarray}
T_{oh}&=&\frac{N_{oh}}{f_{c}}
\\
T_{ack}&=&\frac{N_{ack}}{f_{c}}
\\
T_{data}&=&\frac{N_{data}}{f_{c}}
\end{eqnarray}

Following notations are used
\\\\
$T_{data}=Time$ $for$ $data$ $to$ $reach$ $end$ $of$ $frame$
\\
$T_{ta}=Turnaround$ $Time$
\\
$T_{ack}=Acknowledgement$ $time$
\\
$T_{oh}=OverHead$ $time$
\\
$T_{g}=Guard$ $time$
\\
$f_{c}= Communication$ $Data$ $Rate$
\\
$N_{oh}=Total$ $overhead$ $bits$
\\
$N_{ack}=ACK/NACK$ $message$ $bits$
\\
$N_{data}=Total$ $data$ $bits$
\\
\subsubsection{Throughput of CSMA/CA}

Throughput of CSMA/CA is calculated by formula given in equation 16. Delay in equation 30 is calculated by adding delays of all elements of frame while it reaches receiver.

\begin{eqnarray}
D=T_{bo}+T_{data}+T_{ta}+T_{ack}+T_{ifs}+T_{rts}+T_{cts}
\end{eqnarray}
\\
The following notations are used
\\\\
$T_{bo}=Back$ $Off$ $Period$
\\
$T_{rts}=Request$ $To$ $Send$
\\
$T_{cts}=Clear$ $To$ $Send$
\\
$T_{data}=Transmission$ $Time$ $of$ $Data$
\\
$T_{ta}=Turn$ $Around$ $Time$
\\
$T_{ack}=Acknowledgement$ $Transmission$ $Time$
\\
$T_{ifs}=Inter$ $Frame$ $Space$
\\\\
Now we will calculate delay time given in equation 30
\begin{eqnarray}
T_{bo}=bo_{slots}.T_{boslots}
\\
T_{ta}=T_{data} + T_{ack}
\end{eqnarray}
\\
$bo_{slots}=Back$ $off$ $slots$ $number$
\\
$T_{boslots}=Back$ $off$ $slots$ $time$
\\
\begin{eqnarray}
T_{ack}&=&\frac{N_{ack}}{f_{c}}
\\
T_{ifs}&=&T_{data}-T_{ack}
\end{eqnarray}
Following notations are used
\\\\
$T_{ta}=Turnaround$ $Time$
\\
$T_{ack}=Acknowledgement$ $time$
\\
$f_{c}= Communication$ $Data$ $Rate$
\\
$N_{ack}=ACK/NACK$ $message$ $bits$
\\\\
\indent If there is no acknowledgement then turnaround time $T_{turnaround}$ and $T_{ack}$ is equal to zero.
\\
\section{Simulation Results}
Table III summarizes the simulation parameters with their values used in access techniques. And table IV elaborate each simulation parameter.
\begin{table*}
\caption{Simulation Parameter For Computation Of Multiple Access Techniques}
\begin{center}
\begin{tabular}{ | p{3cm} | p{1.5cm}| p{1.5cm}|p{1.5cm}|p{1.5cm}|p{1.5cm}|p{1.5cm}|p{1.5cm}|}
    \hline
    &\multicolumn {7}{c}{Simulation Parameters}\\ \cline{2-8}

     Name Of Technique Used & N & L(bits) & C(Kbit/s) & tau(msec)& P & $\lambda$ & K\\ \hline
     FDMA   & 100&256&64&5&2e-4	&2&  \\ \hline
     TDMA	& 100&256&64&5&	&&2  \\ \hline
     CSMA   & 100&256&64&5&  &&2  \\ \hline
     ALOHA  & 100&256&64&5&  &&  \\ \hline
     S ALOHA& 100&256&64&5&2e-4 &2&  \\ \hline
     \end{tabular}
     \end{center}
\end{table*}

\subsection{Comparison Of Throughput As a Function Of Delay}
Different multiple access techniques have been compared. Each technique has number of varying parameters. By varying some of the parameters relation between throughout and normalized delay is affected.
\\
\indent N is the number of nodes through which data was sent from transmitter to receiver. L is length of frame and is kept constant for all the techniques. It is kept constant so that the path for sending the data remains same and performance for these techniques are found out. $\lambda$ is the packet arrival rate, rate at which packets are arriving from transmitter to receiver. P is the probability of failure the probability that packet has been not successfully transmitted. Tau is the slot duration. Simulations are carried out by keeping number of nodes, length of frame, frequency digit and slot duration same for all the techniques.
\\
\indent The results are shown in the Fig 9. It represents average normalized delay of TDMA, FDMA, S-ALOHA, Pure ALOHA, CSMA/CA as a function of throughput. Throughput is the successful transfer rate through the medium and average normalized delay is the delays of entire frame from first packet send from transmitter to last packet received at the receiver. TDMA out performs everyone.
\\
\indent CSMA/CA is a carrier sensing technique due to which delay is slightly high, where as TDMA is time division technique and the medium is divided in time slots, each transmitter is sending packets at its own time so there is less delay compare to others. One of the main reason for CSMA/CA having the maximum delay is that it is continuously sensing the medium and waiting for medium to get free and if it finds the medium free then there will be transmission if the medium is not free then CSMA/CA will be keep waiting and will have a long delay for packet transmission. FDMA is the closest technique to TDMA. There is very small difference between FDMA and TDMA, reason is that FDMA is based on division of frequency bands into number of frequencies which transmits data in its own frequency so the delay is minimized as everyone is transmitting in its own band.
\\
\indent Pure ALOHA and S-ALOHA are the two techniques that are closer to each other and also have very less difference in delay compare to CSMA/CA. The reason is that ALOHA sends the data without sensing the medium and as result collision occurs and delay time of sending of data increases. In S-ALOHA the transmission is done at the beginning of the frame and if there is a delay in the sending of data in beginning of frame then the data cannot be send in the middle or end of frame it has to wait for next frame as a result there is a delay.
\\
\indent Calculations have been done by keeping the throughput range from $0.01$ $to$ $10^0$ along the X-Axis and the delay range from $0.01$ to $10^2$ along Y-Axis. Legend shows different lines code which represents different techniques of multiple access. At throughput S=$0.1$ TDMA has an average normalized delay of $0.01$, CSMA/CA has an average normalized delay of $10^2$, FDMA having delay of $0.1$, Slotted ALOHA having normalized delay of $10^{0.5}$ and Pure ALOHA has a delay of $10^{0.7}$. Keeping TDMA throughput as reference, FDMA is closest to TDMA having a difference in delay of $10^{0.2}$, CSMA having a delay difference of $10^4$ where as compared to TDMA, S-ALOHA and Pure ALOHA have a delay difference of $10^{2.7}$ and $10^{2.5}$ respectively. If we keep delay as reference then at an average normalized delay of 7 $\mu$s, multiple access techniques have following throughput TDMA $10^0$, FDMA $10^{-0.05}$, S-ALOHA $10^{-0.7}$, Pure ALOHA $10^{-0.8}$ and CSMA/CA $10^{-1.8}$.
\\
\indent These values show that TDMA at delay of 7 $\mu$s have highest throughput and CSMA/CA have lowest. TDMA outperforms other techniques. Closest to TDMA is FDMA there is very less difference between them. Pure ALOHA and S-ALOHA are closer to each other having a difference of $10^{-0.1}$ between them. When throughput was kept as reference TDMA performed better than all and now when delay was kept as reference TDMA again performed better than other, FDMA was in close distance to TDMA but Pure ALOHA, S-ALOHA and CSMA/CA were away from performance of TDMA as shown in Fig 9.

\begin{figure}[!h]
\centering
\caption{Throughput As a Function Of Delay}
\includegraphics[width=4 in, height=3 in]{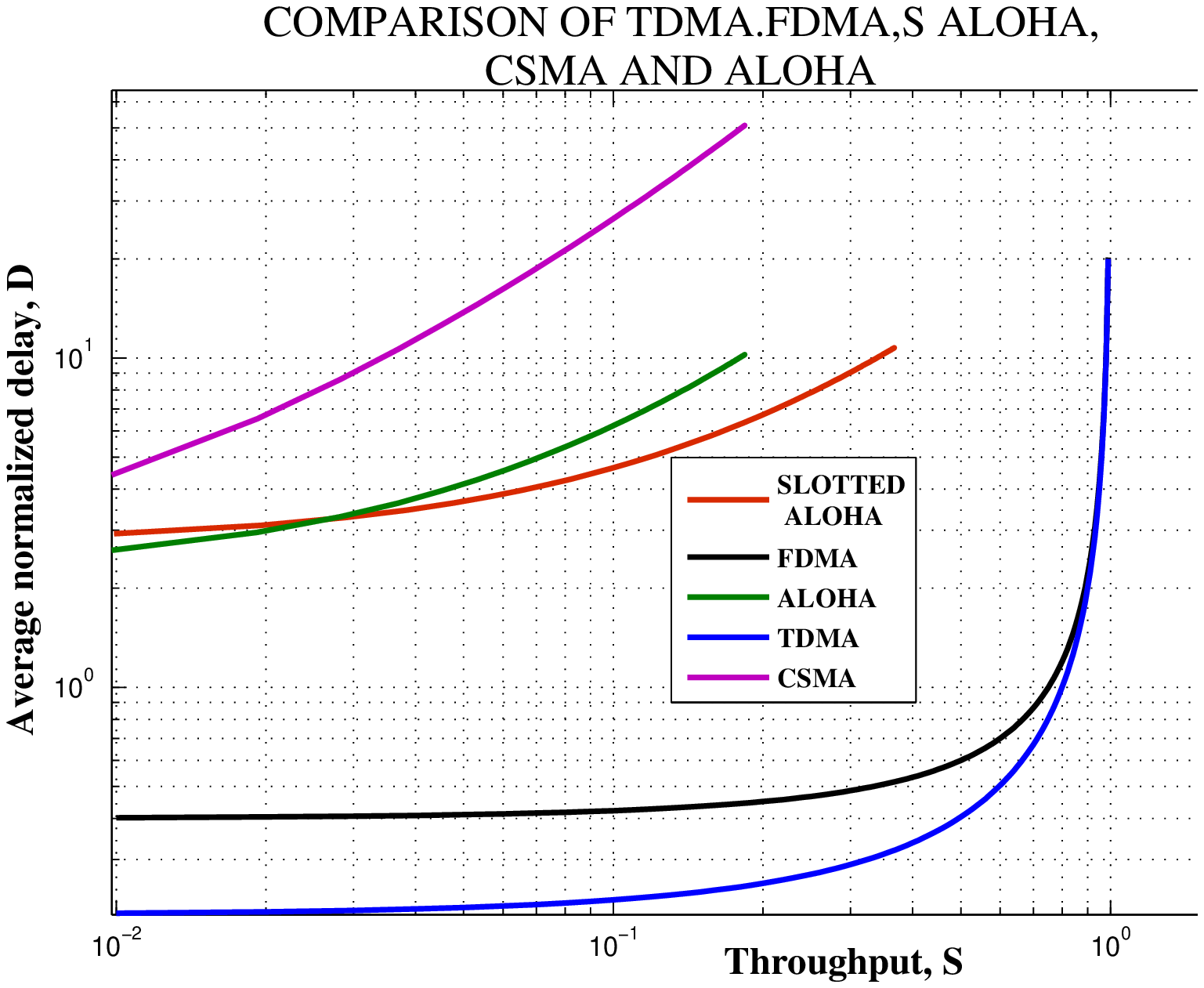}
\end{figure}

\subsection{Comparison of Offered Load as a function of Average Normalized Delay}
In this section, we discuss the effect of offered load on average normalized delay. Each technique has number of varying parameters. By altering some of parameters relationship between offered load and average normalized delay is derived in these techniques. Table III includes parameters which are used in our simulation. Fig 10 depicts the comparison of average normalized delay and offered load of TDMA, FDMA, CSMA/CA, Pure ALOHA and Slotted ALOHA. These protocols are evaluated as function of offered load and average normalized delay.
\\
\indent Offered load is the total traffic load which is offered to a network. Offered load is also defined as traffic generated by nodes in a network. Average normalized delay is the delay of entire frame from first packet sent from transmitter to last packet received at the receiver. CSMA/CA out performs everyone because it has lowest delay as offered load increases. Both offered load and average normalized delay parameters show how well an access protocol performs.

CSMA/CA has a constant delay for increasing traffic load because each node experiences same delay. The closest competitor to CSMA/CA is Slotted ALOHA. In Slotted ALOHA transmission is done at the beginning of frame as there is a delay in the sending of data in beginning of frame then data cannot be send in the middle or end of frame it has to wait for other frame as a result huge delay for increasing load.

TDMA and FDMA have poor performance. As offered load increases delay in TDMA and FDMA becomes unmanageable and same is case with ALOHA and Slotted ALOHA. In FDMA, each node is assigned a different frequency band and node transmits on its particular frequency.When offered load increases delay in FDMA becomes undesirably huge. In TDMA each node is assigned a particular time slot to transmits its data. So as offered load increases delay in TDMA increases as well.

\begin{table}
\caption{Description of Simulation Parameters}
\begin{center}
    \begin{tabular}{ | p{2.5cm} | p{2.5cm} |}
    \hline
    Parameters & Description\\ \hline
    tau(milliseconds) & Slot duration\\ \hline
    K & Kappa\\ \hline
    C(Kbits/s) & Frequency digit\\ \hline
    L(bits) & Length of Frame\\ \hline
    N & Number of nodes\\ \hline
    P & Probability of error\\ \hline
    \end{tabular}
\end{center}
\end{table}
\begin{figure}[!h]
\centering
\caption{Delay and Offered Load}
\includegraphics[width=6 in, height=3 in]{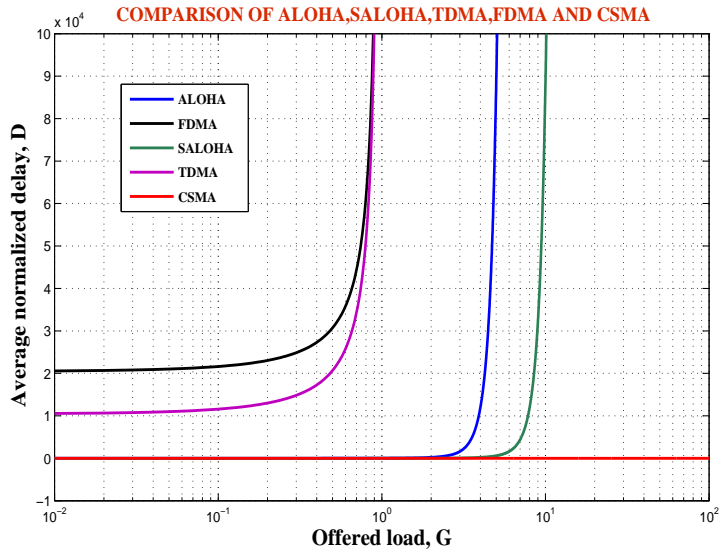}
\end{figure}

\subsection{Comparison of Throughput as a function Offered Load}

We evaluate the effect of offered load on throughput of different medium access protocols. Throughput and offered load are significant parameters for evaluating the performance of accessing techniques. Throughput and offered load performance of an accessing technique shows the capability of handling network resource with increasing capacity in a network.

\begin{figure}[!h]
\centering
\caption{Throughput and Offered Load}
\includegraphics[width=5 in, height=3.2 in]{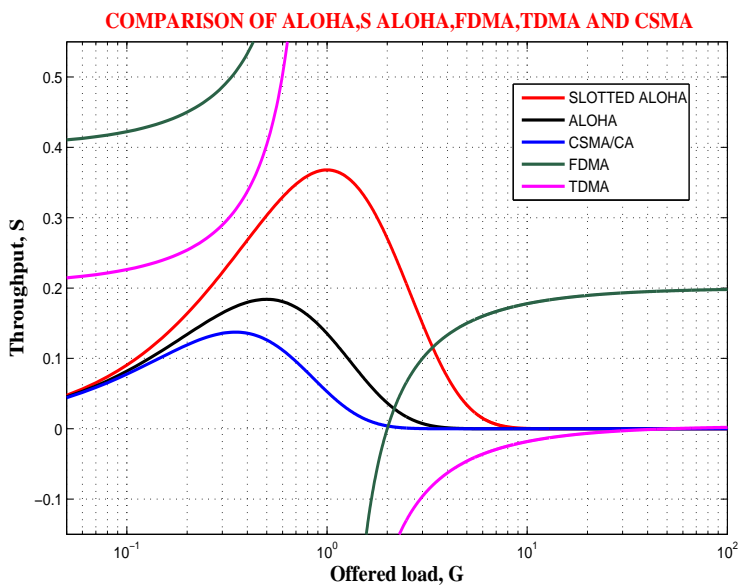}
\end{figure}

Fig 11 depicts the comparison of throughput and offered load of Pure ALOHA, Slotted ALOHA, FDMA, TDMA and CSMA/CA medium access protocols. These protocols are evaluated as function of offered load.

Offered load is total traffic load which is offered to a network. Throughput is the average rate of successful delivery of packet on a communication channel. Offered load and throughput pays pivot role in determining the efficient protocol under increasing traffic.
From Fig 10 it can be deduce that FDMA surpass other protocols. Due to the fact that in FDMA users are assigned different frequencies to access medium and which means each node has different frequency to its packet. So scalability is not an issue in FDMA. TDMA and Slotted ALOHA performance is closest to FDMA. TDMA is a time division technique. Each user is assigned different time slots to access the medium and so scalability is a major problem with TDMA. In Slotted ALOHA transmission can only be initiated at the beginning of frame if data not sent at the beginning of frame then data cannot be send in middle or end of frame.

CSMA/CA and ALOHA have poor performance. CSMA/CA is a contention base protocol. CSMA/CA sense medium before transmitting data onto medium if it found medium to be free then it transmits. In CSMA/CA as offered load increases  collision between packets also increase which indeed results in low throughput. ALOHA sends data without sensing the medium so its throughput is slightly better than CSMA/CA.

For very-low offered load FDMA and TDMA has higher throughput however as offered load increases their throughput gradually becomes constant. Meanwhile Slotted ALOHA performs better in medium-offered load as compared to rest of protocols. However, when offered load increased it performs similar to TDMA, ALOHA and CSMA/CA. So overall Frequency Division Multiple Access (FDMA) outperforms every other protocol.

\section{Conclusion and Future Work}
In this paper different Multiple Access Techniques of MAC protocol which are used in Wireless Body Area Networks have been compared. Techniques are
TDMA, FDMA, CSMA/CA, ALOHA and SALOHA. Algoritham for all these techniques are given in this paper showing their working. Mathematical equations for the calculation
of throughput for all these technqiues have been shown in section IV. Table I shows the comparison of these techniques with different parameters. \\
\indent Performance metrices for the comparison of these techniques are Throughput, Delay and Offered Load. Comparison has been done between performance metirces Throughput and Delay, Delay and Offered Load and Offered Load and Throughput. MATLAB is the tool that is used for simulations. Developing the five access techniques in different scenarios and comparing their graphical results proved that TDMA is the best technqiue to be used in WBAN with increase in load, because it has the highest throught and minimum delay which is the most important requirement of Wireless Body Area Networks. Future work includes performance comparison of these techniques with varying conditions and introducing other metrices.

\end{document}